\documentclass[preprint,aps,12pt,nofootinbib, onecolumn,superscriptaddress,preprintnumbers,balancelastpage,longbibliography]{revtex4-1}

\usepackage{dcolumn} 
\usepackage{xcolor}
\usepackage{youngtab}
\usepackage{ytableau}

\usepackage{booktabs}    
\usepackage{placeins}
\usepackage{soul}

\definecolor{redd}{rgb}{0.8, 0.1,0.2}
\definecolor{navy}{rgb}{0.05, 0.23,0.75}
\usepackage[colorlinks=true,citecolor=red,linkcolor=blue]{hyperref}
\usepackage{chngcntr}
\hypersetup{
     colorlinks   = true,
     citecolor    = navy,
	linkcolor = redd,
	urlcolor=navy,
	anchorcolor=blue
}
\usepackage{bbm}
\usepackage{amsfonts}
\usepackage{amsmath,amssymb}
\usepackage{mathrsfs}
\usepackage{epsfig}
\usepackage{graphicx}
\usepackage{wrapfig}                 
\usepackage{url}
\usepackage{hyperref}
\usepackage{float}
\usepackage{color}
\usepackage{multirow}
\usepackage{lipsum}
\usepackage{enumitem}
\usepackage{tikz}
\usepackage{ctable}
\newcolumntype{L}{>{\centering\arraybackslash}m{1.5cm}}

\usepackage{enumitem}
\usepackage{chngcntr}
\usepackage{array} 

\usepackage[most]{tcolorbox}
\tcbuselibrary{theorems}
\tcbset{
  mytheobox/.style={
    colback=gray!10!white,
    colframe=black,
    boxrule=0.4pt,
    arc=1mm,
    fonttitle=\bfseries,
    title={#1},
    before skip=10pt,
    after skip=10pt,
    breakable,
    enhanced,
  }
}

\newcommand{\nn}{\nonumber}

\newcommand{\be}{\begin{equation}}
\newcommand{\ee}{\end{equation}}
\newcommand{\bea}{\begin{eqnarray}}
\newcommand{\eea}{\end{eqnarray}}
\newcommand{\bc}{\begin{center}}
\newcommand{\ec}{\end{center}}

\begin{document}
		
\title{
Spurion Analysis of $\mathbb{Z}_M/\mathbb{Z}_2$ Non-Invertible Selection Rules: Low-Order versus All-Order Zeros
}

\author{Motoo Suzuki}
\email{msuzuki@sissa.it}
\affiliation{SISSA International School for Advanced Studies, Via Bonomea 265, 34136, Trieste, Italy}
\affiliation{INFN, Sezione di Trieste, Via Valerio 2, 34127, Italy}
\affiliation{IFPU, Institute for Fundamental Physics of the Universe, Via Beirut 2, 34014 Trieste, Italy}

\author{Ling-Xiao Xu}
\email{lxu@ictp.it}
\affiliation{Abdus Salam International Centre for Theoretical Physics, Strada Costiera 11, 34151, Trieste, Italy}

\begin{abstract}
Motivated by recent progress in the spurion analysis of non-invertible selection rules (NISRs) arising from near-group fusion algebras, we further generalize the framework to a class of NISRs obtained from $\mathbb{Z}_2$ orbifolding of a $\mathbb{Z}_M$ symmetry, denoted as $\mathbb{Z}_M/\mathbb{Z}_2$. Many structural features are carried over: for instance, our labeling scheme enables systematic tracking of all couplings when constructing composite amplitudes from simpler building blocks at arbitrary loop orders in perturbation theory. Our analysis provides a transparent understanding of both low-order and all-order zeros of couplings under radiative corrections. Furthermore, we examine the fate of low-order zeros when the fusion algebra is not faithfully realized --- a situation not captured by the vanilla argument of ``loop-induced groupification'' --- and formulate a conjecture on the related aspects of particle decoupling and effective theory. Finally, we discuss the low-order versus all-order zeros in Yukawa textures from the perspective of spurion analysis.  

\end{abstract}

\maketitle
\tableofcontents

\section{Introduction}
\label{sec:intro}

Inspired by the recent generalization of global symmetries~\cite{Gaiotto:2014kfa}, many subtle aspects of quantum field theory (QFT) and lattice models have been revisited and refined over the past decade. This effort has led to the unification of a growing number of results within the framework of \emph{generalized global symmetries} (see e.g.~\cite{Cordova:2022ruw, McGreevy:2022oyu, Gomes:2023ahz, Schafer-Nameki:2023jdn, Brennan:2023mmt, Luo:2023ive, Shao:2023gho, Costa:2024wks, Iqbal:2024pee, Davighi:2025iyk} for reviews). Related applications to various aspects of the Standard Model (SM) of particle physics and beyond include e.g.~\cite{Choi:2022jqy, Cordova:2022ieu, Putrov:2023jqi, Choi:2023pdp, Cordova:2022fhg, Cordova:2023her, Cordova:2024ypu, Delgado:2024pcv, Cao:2024lwg, Choi:2022fgx, Yokokura:2022alv, Hidaka:2024kfx, DelZotto:2024ngj, Gagliano:2025oqv, Tong:2017oea, Garcia-Etxebarria:2018ajm, Davighi:2019rcd, Wan:2019gqr, Wang:2021ayd, Wang:2020mra, Alonso:2024pmq, Li:2024nuo, Koren:2024xof, Alonso:2025rkk, Koren:2025utp, Anber:2021upc, Anber:2025gvb, Cordova:2022qtz, Chen:2025buv, Aloni:2024jpb, Garcia-Valdecasas:2024cqn, Anber:2024gis, Dierigl:2024cxm, Craig:2024dnl, Brennan:2023kpw, Hidaka:2019mfm, Hidaka:2020iaz, Hidaka:2020ucc, Hidaka:2020izy, Hidaka:2021mml, Hidaka:2021kkf, Chen:2024tsx, Choi:2025vxr}.

Recently, a class of selection rules imposed by non-invertible fusion algebras has been revived and has attracted growing attention in the particle physics and string phenomenology communities~\cite{Kaidi:2024wio, Heckman:2024obe, Kobayashi:2024yqq, Kobayashi:2024cvp, Funakoshi:2024uvy, Kobayashi:2025znw, Suzuki:2025oov, Suzuki:2025bxg, Liang:2025dkm, Kobayashi:2025ldi, Kobayashi:2025cwx, Nomura:2025sod, Kobayashi:2025lar, Dong:2025jra, Nomura:2025yoa, Chen:2025awz, Okada:2025kfm, Kobayashi:2025thd, Jangid:2025krp, Kobayashi:2025ocp, Kobayashi:2025rpx, Jiang:2025psz, Nomura:2025tvz, Jangid:2025thp} (see also~\cite{Hamidi:1986vh, Font:1988nc, Kobayashi:1995py, Kobayashi:2011cw} for some early seminal works long ago). In this paper, we refer to these as \emph{non-invertible selection rules} (NISRs). 
Although NISRs may appear unconventional compared to familiar selection rules arising from ordinary group laws, they have been proven useful in a variety of concrete particle physics models. Indeed, one may view them as a natural extension of ordinary symmetry-based constraints. (We hope that, in the near future, NISRs will be regarded as a standard and unexotic tool for model building --- see, e.g.~\cite{Suzuki:2025oov} for simple SM extensions illustrating this point.)
More generally, some \emph{universal} features of NISRs under radiative corrections were systematically analyzed in~\cite{Kaidi:2024wio, Heckman:2024obe}, where the central mechanism of \emph{loop-induced groupification}~\cite{Kaidi:2024wio} was introduced. As we will discuss later, this idea suggests that NISRs provide a novel organizing principle for particle interactions --- hence for scattering amplitudes and quantum corrections --- in perturbation theory.

From the perspective of particle physics, quantum corrections are often organized through \emph{spurion analysis}, in which coupling constants in the theory are promoted to (non-propagating) background fields. Their contributions in various processes --- both at tree and loop levels --- are then systematically tracked via their charges under the associated spurious symmetries. A well-known example is provided by quark masses in chiral perturbation theory, where they are treated as spurions charged under chiral symmetry in the construction of the low-energy effective theory. A closely-related notion is the \emph{'t Hooft technical naturalness}~\cite{tHooft:1979rat}, which states that parameters in a QFT are naturally small and stable under radiative corrections if they are protected by symmetries --- specifically, by those that emerge in the limit where the corresponding couplings are switched off. 

To incorporate NISRs into the standard toolbox of particle physicists, the following questions naturally arise: 
\\[0.25cm]
\textit{To what extent can the results implied by the ``loop-induced groupification'' of NISRs be understood using spurion analysis? How are they related to the conventional notion of 't Hooft naturalness?}
\\[0.4cm]
This question was partially addressed in~\cite{Suzuki:2025bxg} for NISRs arising from near-group fusion algebras~\cite{Evans:2012ta}. In this work, we extend this line of exploration by further generalizing the framework of spurion analysis to the NISRs obtained from $\mathbb{Z}_2$ orbifolding of the $\mathbb{Z}_M$ symmetry, denoted as $\mathbb{Z}_M/\mathbb{Z}_2$. As we will show, many of the structural results established for near-group fusion algebras can be carried over. Notably, when $M=3$ and $4$, the $\mathbb{Z}_M/\mathbb{Z}_2$ fusion algebras are the same as the Fibonacci and Ising fusion algebras, respectively --- both belonging to near-group fusion algebras --- so our results in these cases agree with~\cite{Suzuki:2025bxg}. 

The remainder of the paper is organized as follows.
In Section~\ref{sec:algebra_review}, we begin with a concise review of the $\mathbb{Z}_{M}/\mathbb{Z}_2$ fusion algebra and the argument of “loop-induced groupification”, followed by some original remarks on situations where this mechanism is obstructed.
In Section~\ref{sec:spurion-analysis}, we present the general framework of the spurion analysis and verify its consistency.
Section~\ref{sec:low_order_zeros_decoupling} illustrates this framework with a concrete example and formulates a conjecture concerning unfaithfully-realized fusion algebras and their implications for effective theories.
In Section~\ref{sec:zeros}, we discuss the implications of the spurion analysis for Yukawa texture zeros.
Finally, we provide our concluding remarks in Section~\ref{sec:conclusion}, and present an alternative labeling scheme using basis elements in certain special cases in Appendix~\ref{app:A}.

\section{The $\mathbb{Z}_{M}/\mathbb{Z}_2$ fusion algebra, groupification, and its obstructions}
\label{sec:algebra_review}

We start with a lightning review of the $\mathbb{Z}_{M}/\mathbb{Z}_2$ fusion algebra. (More precisely, its basis elements are the $\mathbb{Z}_2$-invariant conjugacy classes of $D_{M}\equiv \mathbb{Z}_M\rtimes \mathbb{Z}_2$; hence it is sometimes described as the $\mathbb{Z}_2$ gauging of $D_{M}$.) 
We also refer the reader to e.g.~\cite{Kaidi:2024wio, Kobayashi:2024cvp, Funakoshi:2024uvy, Liang:2025dkm} for related discussions.

\begin{figure}[t]
\centering
\includegraphics[scale=0.12]{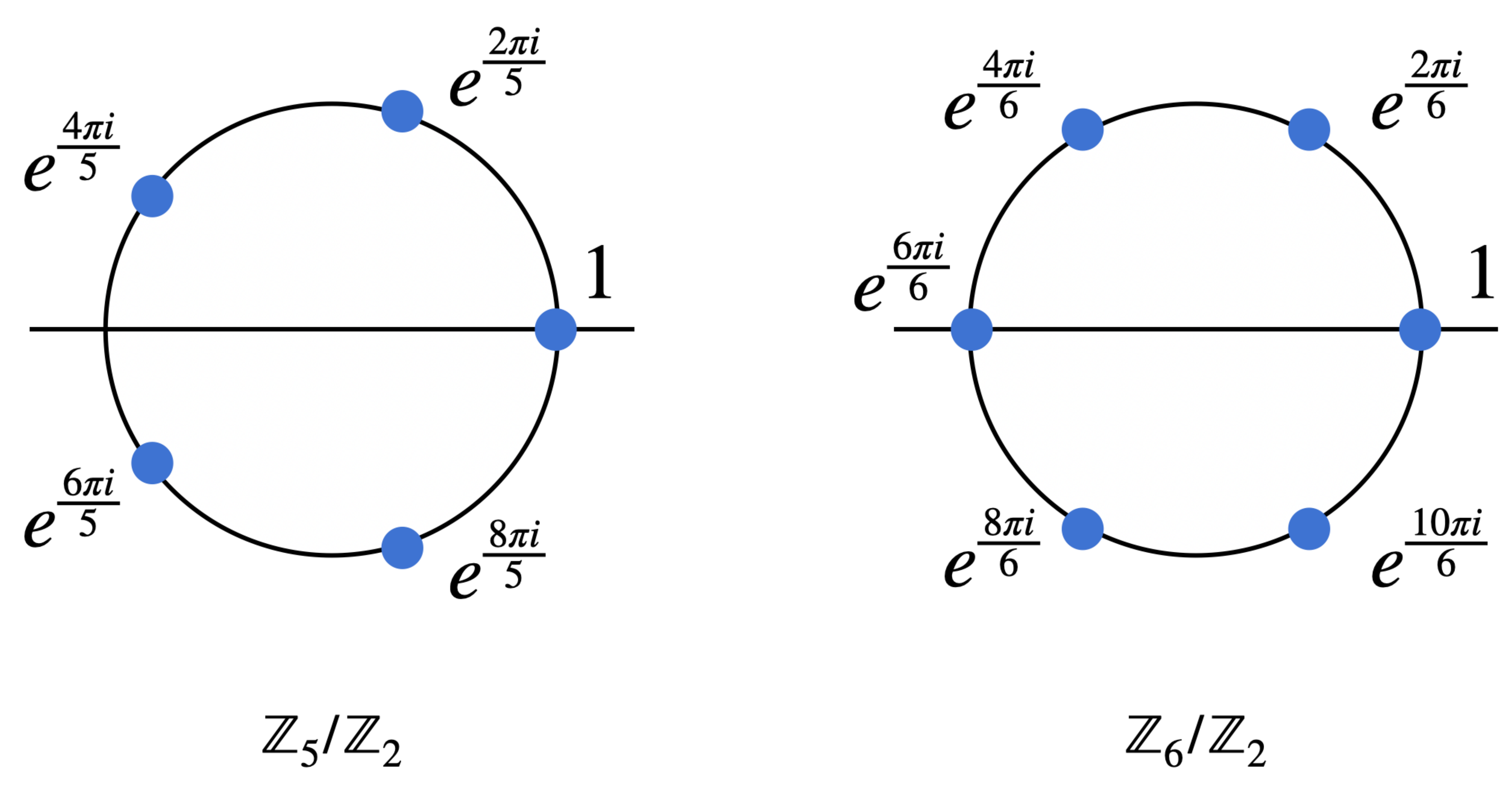}
\caption{An intuitive illustration of the basis elements of $\mathbb{Z}_M/\mathbb{Z}_2$ fusion algebra when $M=5$ and $6$, where each class includes a pair of $\mathbb{Z}_M$ elements related by inversion that are identified by the $\mathbb{Z}_2$ orbifolding.} 
\label{fig:algebra}
\end{figure}

We consider a set of basis elements, each denoted by a \emph{class} $[g^j]$, obtained by $\mathbb{Z}_2$ orbifolding of a $\mathbb{Z}_{M}$ group. Here, $g$ is the generator of a $\mathbb{Z}_M$ group (i.e., the group element $e^{\frac{2\pi i}{M}}$). The classes $[g^j]$ are defined as follows:
\begin{itemize}
\item For $j=0$ with any $M$ and $j=\frac{M}{2}$ with even $M$, the class $[g^j]$ contains a single element: $[g^0]=\{1\}$ and $[g^{\frac{M}{2}}]=\{-1\}$.
\item For all other integers $j$, the class $[g^j]$ contains the two elements of $\mathbb{Z}_{M}$ exchanged by inversion: $[g^j]=\{e^{\frac{2\pi i}{M}j}, e^{-\frac{2\pi i}{M}j}\}$.
\end{itemize}
Intuitively, the elements that are distinct in $\mathbb{Z}_M$ become identified under the $\mathbb{Z}_2$ orbifold, so each class $[g^j]$ includes a pair related by inversion; for instance, see in Fig.~\ref{fig:algebra} for further illustration of the elements in the $\mathbb{Z}_{5}/\mathbb{Z}_2$ and $\mathbb{Z}_{6}/\mathbb{Z}_2$ fusion algebras. From the definition, it follows that 
\be
[g^j] \sim [g^{M-j}]\;,
\ee
since $e^{\frac{2\pi i}{M}j}=e^{-\frac{2\pi i}{M}(M-j)}$ and $e^{-\frac{2\pi i}{M}j}=e^{\frac{2\pi i}{M}(M-j)}$. Consequently, without loss of generality, one may restrict to
\be
j\in \left[0, \frac{M}{2}\right] \quad \text{for even } \quad M, \quad\quad\quad j\in \left[0, \frac{M-1}{2}\right] \quad \text{for odd} \quad M\;.
\ee

The $\mathbb{Z}_{M}/\mathbb{Z}_2$ fusion algebra is defined as follows. For generic integers $j, k$, the fusion rules are
\be
[g^j]\cdot [g^k]= [g^{j+k}] \oplus [g^{j-k}]\; ;
\label{eq:fusion_rules_ZM_mod_Z2_1}
\ee
while for special values  $\frac{M}{2}$ (for even $M$) and $k=0$, the fusion rules are
\be
[g^j] \cdot [g^{\frac{M}{2}}] =[g^{\frac{M}{2}-j}]\;, \quad\quad [g^j] \cdot [g^0] =[g^j]\; .
\label{eq:fusion_rules_ZM_mod_Z2_2}
\ee
(As one can recognize, these fusion rules descend from the ordinary $\mathbb{Z}_M$ group law for each representative in the class $[g^j]$, while the non-invertible nature of the fusion algebra originates from orbifolding.)
From Eqs.~\eqref{eq:fusion_rules_ZM_mod_Z2_1} and~\eqref{eq:fusion_rules_ZM_mod_Z2_2} we immediately see the following features:
\begin{enumerate}
\item Every class $[g^j]$ is \emph{self-conjugate}, i.e. $[g^0]\prec [g^j] \cdot [g^j]$ with $[g^0]$ being the identity. Except for the special cases $j=0, \frac{M}{2}$ with even $M$ and $j=0$ with odd $M$, $[g^j]$ is \emph{non-invertible} since $[g^j] \cdot [g^j] \neq [g^0]$.
\item The fusion algebra is clearly commutative: $[g^j]\cdot [g^k]=[g^k]\cdot [g^j]$.
\end{enumerate}

The $\mathbb{Z}_{M}/\mathbb{Z}_2$ fusion algebra can be used as NISRs to construct particle physics models in perturbation theory. Specifically, each \emph{dynamical} particle $\phi_i$ in the theory is labeled by a basis element $[g^{j_i}]$ in the $\mathbb{Z}_{M}/\mathbb{Z}_2$ fusion algebra, i.e., 
\be
\ell(\phi_i) = [g^{j_i}]\;,
\ee
and a term is allowed in the \emph{classical} Lagrangian $\mathcal{L}_{cl}$ when the fusion product of the corresponding labels contains the identity element~\footnote{Note that the classical Lagrangian in Eq.~\eqref{eq:classical_Lag_exp} corresponds to interactions between scalars without additional quantum numbers. Otherwise, some interactions allowed by the $\mathbb{Z}_{M}/\mathbb{Z}_2$ fusion algebra may be forbidden by additional quantum numbers.}, i.e., 
\be
\mathcal{L}_{cl} \supset \lambda_{\phi_1 \phi_2 \cdots \phi_N} \phi_1 \phi_2 \cdots \phi_N \quad \text{when} \quad [g^0]\prec [g^{j_1}] \cdot  [g^{j_2}] \cdot ... \cdot [g^{j_N}]\;.
\label{eq:classical_Lag_exp}
\ee
Here $\lambda_{\phi_1 \phi_2 \cdots \phi_N}$ denotes the coupling that will be viewed as a non-propagating \emph{background} field in spurion analysis.~\footnote{Since we only consider Lorentz-invariant theories, those background fields do not carry spacetime indices.} Each term in $\mathcal{L}_{cl}$ can thus be viewed as a contact, tree-level amplitude involving both the background and dynamical fields. Since every $[g^j]$ is self-conjugate, kinetic terms are allowed for all dynamical particles.

The classical Lagrangian encodes only the tree-level vertices; additional terms arise at the loop level from quantum fluctuations. In particular, \emph{some} interactions forbidden at the tree level can be radiatively generated. As a result, the NISRs imposed by the $\mathbb{Z}_{M}/\mathbb{Z}_2$ fusion algebra --- while exact at the tree level --- get increasingly violated at higher loop orders and eventually reduce to a finite Abelian group. This process of radiative violation of NISRs implies both \emph{low-order zeros} and \emph{all-order zeros} for couplings that can appear at different loop orders in perturbation theory. 
Formally, this is systematically understood as ``loop-induced groupification''~\cite{Kaidi:2024wio}: once one allows dressings by non-identity elements produced in the fusions of conjugate pairs $([g^l])^2$, distinct classes $[g^j]$ and $[g^k]$ become \emph{indistinguishable and equivalent} at the loop level.

\begin{figure}[t]
\centering
\includegraphics[scale=0.2]{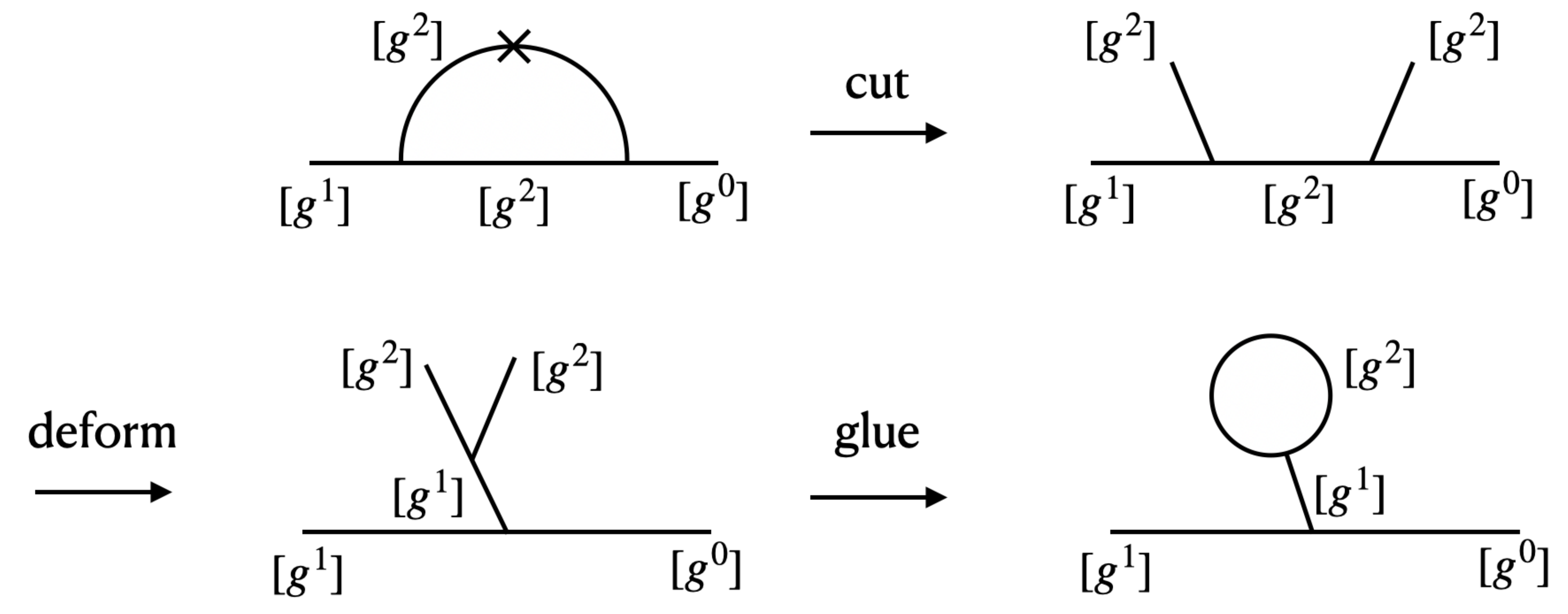}
\caption{An example of groupification for the $\mathbb{Z}_5/\mathbb{Z}_2$ fusion algebra. Due to quantum corrections, the particle labeled by the non-invertible element $[g^2]$ appears in the loop and induces the identification (i.e., mixing) between the particles labeled by $[g^1]$ and $[g^0]$ at the one-loop level. By deforming the diagrams, we see that $[g^1]$ and $[g^0]$ become identified by dressing $[g^1]$ produced by the fusion of conjugate pair $([g^2])^2$. This matches Eq.~\eqref{eq:groupification}. The same mechanism applies to other fusion algebras; see e.g.~\cite{Suzuki:2025oov}.} 
\label{fig:groupification}
\end{figure}

Some comments are in order:
\begin{itemize}
\item Assuming the $\mathbb{Z}_{M}/\mathbb{Z}_2$ fusion algebra is realized faithfully --- i.e., each $[g^j]$ labels at least one dynamical particle in the theory --- then $[g^j]$ and $[g^k]$ become indistinguishable at the one-loop order by dressing a conjugate pair $([g^l])^2$ when 
\be
[g^j]\prec [g^k] ([g^l])^2\;.
\label{eq:groupification}
\ee
This equation means that $[g^j]$ exists in the fusion product of $[g^k] ([g^l])^2$.
At higher loop orders, a similar equation holds by dressing more conjugate pairs.
Following the intuitive diagrammatic illustration in~\cite{Suzuki:2025oov}, Eq.~\eqref{eq:groupification} can be exemplified in Fig.~\ref{fig:groupification} for $M=5$.~\footnote{The same argument applies to other values of $M$.} 
From Eq.~\eqref{eq:groupification}, ``loop-induced groupification'' implies: 
\begin{enumerate}
\item For odd $M$, the $\mathbb{Z}_{M}/\mathbb{Z}_2$ fusion algebra reduces to the trivial group (i.e., the group which only has the identity) at the one-loop order. We denote the all-order exact group as
\be
\text{Gr}[\mathbb{Z}_M/\mathbb{Z}_2]=\{1\}\quad \text{for odd} \quad M\;. 
\label{eq:Gr_odd}
\ee
\item For even $M$, the fusion rules in Eqs.~\eqref{eq:fusion_rules_ZM_mod_Z2_1} and~\eqref{eq:fusion_rules_ZM_mod_Z2_2} reduce to a $\mathbb{Z}_2$ group that is exact up to all loop orders, where the $[g^j]$ with odd $j$ carry charge $-1$ and those with even $j$ carry charge $+1$. Hence, the all-order exact group is denoted as 
\be
\text{Gr}[\mathbb{Z}_M/\mathbb{Z}_2]=\mathbb{Z}_2 \quad \text{for even} \quad M\;. 
\label{eq:Gr_even}
\ee
\end{enumerate}
The above result is easily seen as follows. For odd $M$, $[g^1]\prec \left([g^{\frac{M-1}{2}}]\right)^2$, so dressing $[g^1]$ as in Eq.~\eqref{eq:groupification} identifies all classes at one loop. 
For even $M$, the class $[g^{\frac{M-1}{2}}]$ does not exist; instead $[g^2]\prec \left([g^{\frac{M}{2}-1}]\right)^2$ or $[g^2]\prec \left([g^1]\right)^2$, which distinguishes classes with even and odd $j$'s, rendering a $\mathbb{Z}_2$ group at all loop orders. The same conclusion was reached in~\cite{Kaidi:2024wio, Funakoshi:2024uvy}.

\item The above result from groupification relies on two key assumptions:
\begin{enumerate}
\item \textbf{Faithful realization:} the theory contains dynamical particles labeled by all classes, so that all $[g^j]$ equivalent to a given $[g^k]$ can be identified via Eq.~\eqref{eq:groupification} by dressing all possible $([g^l])^2$. 
\item \textbf{No additional quantum numbers:} the relevant particles carry no extra quantum numbers that would forbid the required mixings or interactions. 
\end{enumerate}
If either condition fails, the identification in Eq.~\eqref{eq:groupification} --- and hence loop-induced groupification --- can be obstructed: e.g. if particles labeled by some $[g^l]$ are absent~\footnote{As in Fig.~\ref{fig:groupification} for the $\mathbb{Z}_5/\mathbb{Z}_2$ NISRs, if the particle labeled by $[g^2]$ does not exist, the particles labeled by $[g^1]$ and $[g^0]$ cannot be identified at the one-loop level. We will discuss more on this point in Section~\ref{sec:low_order_zeros_decoupling}.}, or if additional symmetries forbid the necessary couplings~\footnote{For example, if $[g^k]$ labels a scalar and $[g^j]$ labels a fermion, Lorentz symmetry forbids their mixing at all loop orders, although they might get identifed in the groupification of the $\mathbb{Z}_M/\mathbb{Z}_2$ fusion algebra.}.
Therefore, caution is advised when applying the argument of ``loop-induced groupification'' to specific particle physics models, where the fusion algebra may be unfaithfully realized and particles often carry additional quantum numbers.  
\end{itemize}

\section{Spurion analysis for the $\mathbb{Z}_{M}/\mathbb{Z}_2$ NISRs}
\label{sec:spurion-analysis}

In this section, we first present a general discussion of the spurion analysis, followed by a concrete example in Section~\ref{sec:low_order_zeros_decoupling}. The reader may be advised to refer back and forth between the general discussion and the example for a more coherent understanding. 

\subsection{Labeling scheme for couplings}
\label{sec:general_theory}

As shown in~\cite{Suzuki:2025bxg}, in the case of near-group fusion algebras, a consistent labeling of the tree-level couplings in the classical Lagrangian allows one to reconcile the mechanism of loop-induced groupification of NISRs with the conventional spurion analysis. This labeling scheme enables a systematic tracking of coupling contributions to scattering processes at arbitrary loop orders in perturbation theory. 
The key insight is that some tree-level couplings should be labeled by \emph{non-identity} elements, even when NISRs are exact at the tree level.
In this work, we generalize the framework of spurion analysis to the $\mathbb{Z}_M/\mathbb{Z}_2$ fusion algebras, which are not the near-group fusions for $M\geq 5$. (See~\cite{Evans:2012ta} for the definition of near-group fusions.)

One convenient way to label the coupling constants is through the use of a \emph{lifted group}~\cite{Suzuki:2025bxg} --- the group that is identified when all tree-level couplings with non-identity (i.e., nontrivial) labels are switched off. In this view, NISRs can be interpreted as a specific way of breaking the lifted group, thereby justifying that the couplings with nontrivial labeling are technically natural in the sense of 't Hooft~\cite{tHooft:1979rat}. 

For the $\mathbb{Z}_M/\mathbb{Z}_2$ fusion algebras, all elements are self-conjugate~\footnote{This is analogous to the non-invertible element $\rho$ in the near-group fusions; see~\cite{Suzuki:2025bxg}.}, making it natural to use a $\mathbb{Z}_2$ group --- which we denote as $(\mathbb{Z}_2)_{\text{lift}}$ --- for each basis element of the fusion algebra.~\footnote{Since the identity element does not play any role in the spurion analysis, one may restrict only to the non-identity elements.}
The $(\mathbb{Z}_2)_{\text{lift}}$ group captures whether a given basis element appears as conjugate pairs in a scattering process. Specifically, we introduce a $\mathbb{Z}_2$-valued function $X([g^j];\mathcal{M})$ that counts the number of unpaired particles labeled by $[g^j]$ in a scattering process $\mathcal{M}$: $X([g^j];\mathcal{M})=0$ and $1$ when there are even and odd number of external dynamical particles labeled by $[g^j]$, respectively. As noticed in~\cite{Suzuki:2025bxg}, $X([g^j];\mathcal{M})$ satisfies
\be
X([g^j]; \mathcal{M}_1^{(0)}\cup_c \mathcal{M}_2^{(0)}) = X([g^j]; \mathcal{M}_1^{(0)}) + X([g^j]; \mathcal{M}_2^{(0)}) \quad \text{mod} \quad 2,
\label{eq:X}
\ee
where $\mathcal{M}_1^{(0)}$ and $\mathcal{M}_2^{(0)}$ are two arbitrary tree scattering processes (the superscript indicating their loop order), and $\mathcal{M}_1^{(0)}\cup \mathcal{M}_2^{(0)}$ denotes the composition obtained by gluing a conjugate pair of dynamical particles labeled by $c$.~\footnote{In the following, we suppress the label $c$, since our argument does not rely on which conjugate pair is glued in the construction.} 
Clearly, Eq.~\eqref{eq:X} is consistent with the ordinary group law of $(\mathbb{Z}_2)_{\text{lift}}$: we assign $(\mathbb{Z}_2)_{\text{lift}}$-charge $+1$ for $\mathcal{M}$ when $X([g^j];\mathcal{M})=0$ and charge $-1$ when $X([g^j];\mathcal{M})=1$, respectively.
Consequently, all scattering processes can be classified according to their charges under the full lifted group
\be
G_{\text{lift}} = \mathbb{Z}_2 \times \mathbb{Z}_2 \times \cdots \times \mathbb{Z}_2, 
\label{eq:lifted_G}
\ee
which encodes whether each $[g^j]$ associated with the external dynamical particles appears in conjugate pairs or not. The number of $\mathbb{Z}_2$ factors in $G_{\text{lift}}$ match the number of (non-identity) elements in the $\mathbb{Z}_M/\mathbb{Z}_2$ fusion algebra.

As noted after Eq.~\eqref{eq:classical_Lag_exp}, each term in the classical Lagrangian can be viewed as a contact amplitude at the tree-level involving both dynamical and background external fields. They serve as the fundamental building blocks from which more complicated scattering processes --- at both tree and loop level --- are constructed by gluing together dynamical particles, i.e., propagating particles. The external legs of the resulting amplitudes then consist of the remaining unglued dynamical fields together with all background fields originating from their simpler sub-amplitudes.
The goal of the spurion analysis is to consistently track the contributions of the couplings in Eq.~\eqref{eq:classical_Lag_exp} across all scattering processes. For this purpose, we propose the following labeling scheme for the coupling of the scattering process $\mathcal{M}$:
\begin{itemize}
\item \label{rule}\textit{All paired labels --- i.e., the $[g^j]$ satisfying $X([g^j];\mathcal{M})=0$ --- for external dynamical particles are replaced by the identity. The overall coupling of $\mathcal{M}$ is then labeled by the fusion product of the remaining unpaired labels, namely those with $X([g^j];\mathcal{M})=1$.}
\end{itemize}
In general, the resulting label of the coupling determined as above is not itself a basis element of the fusion algebra, but rather a fusion product of such elements. We therefore refer to these couplings as \emph{composite spurions}. As one can see, only the couplings associated with vertices whose external particles are all paired are labeled by the identity; accordingly, the couplings are singlets (i.e., uncharged) under $G_{\text{lift}}$. 

\textbf{Example.} Consider a vertex 
\be
\lambda_{[g^1] [g^2]^2 [g^3]} \  [g^1] [g^2]^2 [g^3]
\ee
in the $\mathbb{Z}_M/\mathbb{Z}_2$ fusion algebra for any $M$. Such a vertex is allowed at the tree level, and its coupling is labeled as 
\be
\ell\left(\lambda_{[g^1] [g^2]^2 [g^3]}\right)= [g^1] [g^3]\;. 
\ee
Likewise, a term 
\be
\lambda_{[g^1] [g^3]} \  [g^1] [g^3]
\ee
is forbidden in the classical Lagrangian, but its coupling has the same label, i.e., 
\be
\ell\left(\lambda_{[g^1] [g^3]}\right)= [g^1] [g^3]\;. 
\ee
Two comments are in order:
\begin{enumerate}
\item Indeed, the latter vertex can be induced radiatively from the former by gluing together the dynamical particle $[g^2]$. This justifies why paired labels for external dynamical particles should be replaced by the identity. (As a result, it implies that scattering processes that differ only by conjugate-paired external particles have couplings with identical labeling.) 
\item From the perspective of $G_{\text{lift}}$, both the couplings (and hence the vertices) are charged as $(+1, -1, +1, -1, +1, \cdots)$ under the $\mathbb{Z}_2$ factors corresponding to the basis elements $[g^0], [g^1], [g^2], [g^3], [g^4]$, etc.
\end{enumerate}
Below, we will discuss the relations among different scattering processes more systematically.

We note that, since our argument relies only on the property that all basis elements in the fusion algebra are self-conjugate, the same result applies directly to the Conj$(S_3)$ fusion algebra, which was previously studied as an example beyond near-group fusions in~\cite{Suzuki:2025bxg}.
Furthermore, the labeling scheme introduced above coincides with those of the Fibonacci and Ising fusion algebras for $M=3$ and $4$, respectively~\cite{Suzuki:2025bxg}. A feature of the near-group fusion algebras (and Conj$(S_3)$ as well) is that the fusion product of any unpaired elements always yields a basis element. That is why ``composite spurions'' did not appear in~\cite{Suzuki:2025bxg}.

\subsection{Consistency of the labeling scheme}
\label{sec:general_theory_2}

Let us now demonstrate that the labeling scheme proposed in Section~\ref{sec:general_theory} enables one to systematically track the contributions of couplings when constructing composite amplitudes from simpler building blocks at arbitrary loop orders in perturbation theory. Following the inductive proof in~\cite{Suzuki:2025bxg}, we begin with the case with only tree-level amplitudes and then extend the analysis to the loop level. 

Let us consider a general tree-level scattering amplitude $\mathcal{M}^{(0)}$ (where the superscript indicates the loop order), with the external dynamical particles labeled by the basis elements of the $\mathbb{Z}_M/\mathbb{Z}_2$ fusion algebra:
\be
\mathcal{M}^{(0)} = \lambda_{\mathcal{M}^{(0)}} \ [g^0]^{i_0} [g^1]^{i_1} [g^2]^{i_2} \cdots\;,
\label{eq:tree_amp1}
\ee
where $\lambda_{\mathcal{M}^{(0)}}$ denotes the tree coupling, and the indices $i_0, i_1, i_2, \cdots$ indicate the powers of the corresponding basis elements appearing in $\mathcal{M}^{(0)}$. Each term in the classical Lagrangian in Eq.~\eqref{eq:classical_Lag_exp} represents a contact amplitude; conversely, the amplitudes $\mathcal{M}^{(0)}$ in Eq.~\eqref{eq:tree_amp1} need not be contact interactions --- instead, they can also be constructed from the contact amplitudes originating from the classical Lagrangian. Clearly, $X([g^j]; \mathcal{M}^{(0)})=0$ or $1$ when $i_j$ is even or odd, respectively. Following the labeling scheme in Section~\ref{sec:algebra_review}, we have 
\be
\ell\left(\lambda_{\mathcal{M}^{(0)}}\right) = [g^i] [g^j] [g^k] \cdots \;,
\label{eq:tree_amp1_coup}
\ee
where $[g^i], [g^j], [g^k]$ etc are the remaining unpaired labels in $\mathcal{M}^{(0)}$.~\footnote{To avoid confusion, for any odd power $i_j$ of $[g^j]$, we count only one power in the labeling of Eq.~\eqref{eq:tree_amp1_coup}, since the other $i_j-1$ powers of $[g^j]$ form conjugate pairs.}

Two tree-level amplitudes, $\mathcal{M}_1^{(0)}$ and $\mathcal{M}_2^{(0)}$, can be glued together through a conjugate pair of dynamical particles carrying the same label $[g^j]$. (Recall that all basis elements are self-conjugate in the $\mathbb{Z}_M/\mathbb{Z}_2$ fusion algebra.) The resulting amplitude takes the form
\be
\mathcal{M}_1^{(0)}\cup \mathcal{M}_2^{(0)} = \lambda_{\mathcal{M}_1^{(0)}\cup \mathcal{M}_2^{(0)}} \ [g^0]^{i^\prime_0} [g^1]^{i^\prime_1} [g^2]^{i^\prime_2} \cdots\;,
\label{eq:tree_amp2}
\ee
where the coupling is labeled by the unpaired elements:
\be
\ell\left(\lambda_{\mathcal{M}_1^{(0)}\cup \mathcal{M}_2^{(0)}}\right) = [g^{i^\prime}] [g^{j^\prime}] [g^{k^\prime}] \cdots \;.
\label{eq:tree_amp2_coup}
\ee
From Eq.~\eqref{eq:X}, it follows that
\be
\ell\left(\lambda_{\mathcal{M}_1^{(0)}\cup \mathcal{M}_2^{(0)}}\right) \prec \ell\left(\lambda_{\mathcal{M}_1^{(0)}}\right) \ell\left(\lambda_{\mathcal{M}_2^{(0)}}\right)\;.
\label{eq:spurion_tree}
\ee
The consistency of Eq.~\eqref{eq:spurion_tree} can also be verified by tracking the charges of $\mathcal{M}$'s under $G_{\text{lift}}$; see the discussion before Eq.~\eqref{eq:lifted_G}. Eq.~\eqref{eq:spurion_tree} implies that unpaired labels in the composite amplitude $\mathcal{M}_1^{(0)}\cup \mathcal{M}_2^{(0)}$ cannot be generated from the paired ones in the sub-amplitudes $\mathcal{M}_1^{(0)}$ and $\mathcal{M}_2^{(0)}$. For instance, if $\ell\left(\lambda_{\mathcal{M}_1^{(0)}\cup \mathcal{M}_2^{(0)}}\right)$ contains a label $[g^j]$, it must originate from either $\ell\left(\lambda_{\mathcal{M}_1^{(0)}}\right)$ or $\ell\left(\lambda_{\mathcal{M}_2^{(0)}}\right)$.

This completes our discussion at the tree level. In the following, we justify the validity of an equation in the same form as Eq.~\eqref{eq:spurion_tree} for loop-level amplitudes at arbitrary loop order in perturbation theory. 

Let us now consider an amplitude at the loop order $N$, which we denote as
\be
\mathcal{M}^{(N)} = \lambda_{\mathcal{M}^{(N)}} \ [g^0]^{i_0} [g^1]^{i_1} [g^2]^{i_2} \cdots\;,
\ee
where $\lambda_{\mathcal{M}^{(N)}}$ is the overall coupling, and the elements $[g^j]$ denote the external dynamical particles appearing with multiplicities $i_j$.
Such an amplitude can be reduced to a tree-level one by cutting $N$ internal lines properly. The resulting tree-level amplitude contains all the external legs of $\mathcal{M}^{(N)}$ together with another $N$ additional conjugate pairs --- which we denote as $([h_n])^2$ with $n=1, 2, \cdots, N$ --- corresponding to the internal lines being cut, i.e., 
\be
\mathcal{\tilde{M}}^{(0)} = \lambda_{\mathcal{\tilde{M}}^{(0)}} \ [g^0]^{i_0} [g^1]^{i_1} [g^2]^{i_2} \cdots ([h_1])^2 ([h_2])^2 \cdots ([h_N])^2\;.
\ee
Since all the paired elements are replaced by the identity in the labeling scheme, the appearance of these additional conjugate pairs does not affect the labeling of the overall couplings, i.e., 
\be
\ell\left(\lambda_{\mathcal{M}^{(N)}}\right) = \ell\left(\lambda_{\mathcal{\tilde{M}}^{(0)}}\right)\;.
\label{eq:loop-tree}
\ee
Eqs.~\eqref{eq:loop-tree} and~\eqref{eq:spurion_tree} together imply that 
\be
\ell\left(\lambda_{\mathcal{M}^{(N)}}\right) \prec \ell\left(\lambda_{\mathcal{M}_1^{(0)}}\right) \ell\left(\lambda_{\mathcal{M}_2^{(0)}}\right) \cdots\;.
\label{eq:spurion_loop}
\ee
In words, the labeling of the overall coupling for any loop amplitude is contained within the fusion product of the coupling labels of all the building blocks from which it is constructed.~\footnote{A simple example was already considered at the end of last section, where we have $\ell\left(\lambda_{[g^1] [g^3]}\right) \prec \ell\left(\lambda_{[g^1] [g^2]^2 [g^3]}\right)$.}

This completes our formal discussion of the spurion analysis for all scattering processes in perturbation theory.

Before concluding this section, we make the following remarks:
\begin{itemize}
\item The all-order exact $\mathbb{Z}_2$ symmetry implied by the standard argument of loop-induced groupification --- i.e., Eq.~\eqref{eq:Gr_even} for even $M$ --- arises automatically from the fact that, for even $M$, each interaction term in the classical Lagrangian contains in total an \emph{even} number of external dynamical particles that are labeled by $[g^j]$'s with odd
$j$'s~\footnote{This follows from the fusion algebra reviewed in Section~\ref{sec:algebra_review}.}; consequently, so as does the label for the corresponding coupling. If the $\mathbb{Z}_2$ symmetry were explicitly broken, one would necessarily generate a vertex containing in total an \emph{odd} number of external dynamical particles that are labeled by $[g^j]$'s with odd $j$'s, which is not consistent with Eq.~\eqref{eq:spurion_tree} or~\eqref{eq:spurion_loop}, provided the fundamental building blocks from the classical Lagrangian.

\item In contrast, for odd $M$, the all-order exact group is trivial --- i.e., see Eq.~\eqref{eq:Gr_odd} --- following the argument of loop-induced groupification. From spurion analysis, this is because, for odd $M$, at least one interaction term in the classical Lagrangian contains in total an odd number of external dynamical particles labeled by $[g^j]$'s with odd $j$'s. For instance, the vertex $[g^{\frac{M-1}{2}}]^2 [g^1]$ can exist and break the $\mathbb{Z}_2$ group.

\item Our use of composite spurions may seem unconventional at first glance. However, we emphasize that this structure inherently emerges from the lifted group in Eq.~\eqref{eq:lifted_G}, under which the couplings are charged; see the relevant discussion in Section~\ref{sec:general_theory}. The lifted group ensures the technical naturalness of the nontrivially charged (i.e., non-singlet) couplings. We also refer the reader to Eqs.~\eqref{eq:label_example_Z5} and~\eqref{eq:charge_example_Z5} for a concrete example illustrating the correspondence between the labels and the charges.

\item The NISRs from the $\mathbb{Z}_M/\mathbb{Z}_2$ fusion algebra can be interpreted as a special way of breaking the lifted group in Eq.~\eqref{eq:lifted_G}. In particular, the $\mathbb{Z}_M/\mathbb{Z}_2$ fusion algebra enforces a pattern of hierarchical couplings that cannot be explained using only the lifted group, where different couplings are distinguished by their loop orders. For instance, the vertices $[g^1] [g^2]^2 [g^3]$ and $[g^1] [g^3]$ are charged the same under the lifted group~\footnote{By definition, all couplings are uncharged under the all-order exact group implied by groupification. Relatedly, the hierarchical couplings implied by NISRs cannot be explained by the all-order exact group, either.}, but they should appear at different loop orders as enforced by the $\mathbb{Z}_M/\mathbb{Z}_2$ fusion algebra --- i.e., $[g^1] [g^2]^2 [g^3]$ appears at the tree level, while $[g^1] [g^3]$ appears only at the one-loop level.
A similar observation was made for near-group fusion algebras in~\cite{Suzuki:2025oov, Suzuki:2025bxg}.

\item We note that spurion analysis remains applicable regardless of whether the fusion algebra is faithfully realized or the particles carry additional quantum numbers.
\end{itemize}

\section{The $\mathbb{Z}_5/\mathbb{Z}_2$ NISRs as an example and a conjecture on effective theories}
\label{sec:low_order_zeros_decoupling}

In this section, we illustrate the general discussion of spurion analysis using the $\mathbb{Z}_5/\mathbb{Z}_2$ fusion algebra as a concrete example. The same considerations apply to other fusion algebras as well.

The $\mathbb{Z}_5/\mathbb{Z}_2$ fusion algebra contains three basis elements $[g^0], [g^1], [g^2]$, which satisfy the fusion rules shown in Table~\ref{tab:Z5}. Clearly, the elements $[g^1]$ and $[g^2]$ do not have inverses. In the following analysis in this section, we assume that particles are labeled by these basis elements, i.e., 
\be
\ell(\phi_0) = [g^0], \quad \ell(\phi_1) = [g^1], \quad \ell(\phi_2) = [g^2]\;.
\label{eq:exp_particle_labels}
\ee
For simplicity, we neglect all other quantum numbers associated with these particles.

\begin{table}[h]
\centering
\renewcommand{\arraystretch}{1.0}
\setlength{\tabcolsep}{12pt}
\begin{tabular}{c|ccc}
 & $[g^0]$ & $[g^1]$ & $[g^2]$ \\
\hline
$[g^0]$ & $[g^0]$ & $[g^1]$ & $[g^2]$ \\
$[g^1]$ & $[g^1]$ & $[g^0]+[g^2]$ & $[g^1]+[g^2]$ \\
$[g^2]$ & $[g^2]$ & $[g^1]+[g^2]$ & $[g^1]+[g^0]$ \\
\end{tabular}
\caption{The $\mathbb{Z}_5/\mathbb{Z}_2$ fusion algebra with the basis elements denoted as $[g^0]$, $[g^1]$, and $[g^2]$.}
\label{tab:Z5}
\end{table}

When the $\mathbb{Z}_5/\mathbb{Z}_2$ fusion algebra is realized faithfully, one can construct the classical Lagrangian following Eq.~\eqref{eq:classical_Lag_exp}:
\bea
\mathcal{L}_{\mathbb{Z}_5/\mathbb{Z}_2} (\phi_0, \phi_1, \phi_2) &=& \lambda_{0} \ \phi_0 + \lambda_{0^2} \ \phi_0^2 + \lambda_{1^2} \ \phi_1^2 + \lambda_{2^2} \ \phi_2^2 \nn\\
&+& \lambda_{0^3} \ \phi_0^3 + \lambda_{0 1^2} \ \phi_0 \phi_1^2 + \lambda_{0 2^2} \ \phi_0 \phi_2^2 + \lambda_{1 2^2} \ \phi_1 \phi_2^2 + \lambda_{1^2 2} \ \phi_1^2 \phi_2 \nn\\
&+& \lambda_{0^4} \ \phi_0^4  + \lambda_{0^2 1^2} \ \phi_0^2 \phi_1^2 + \lambda_{0^2 2^2} \ \phi_0^2 \phi_2^2 +\lambda_{0 1 2^2} \ \phi_0 \phi_1 \phi_2^2 + \lambda_{0 1^2 2} \ \phi_0 \phi_1^2 \phi_2 \nn\\
&+& \lambda_{1^2 2^2} \ \phi_1^2 \phi_2^2 + \lambda_{1 2^3} \ \phi_1 \phi_2^3 + \lambda_{1^3 2} \ \phi_1^3 \phi_2 + \lambda_{1^4} \ \phi_1^4+\lambda_{2^4} \ \phi_2^4 \;.
\label{eq:Lag_Z5}
\eea
Higher-dimensional operators can also appear, but they are not included explicitly in $\mathcal{L}_{\mathbb{Z}_5/\mathbb{Z}_2}$.
According to the standard argument of loop-induced groupification, all basis elements become identified at the one-loop level. Consequently, the $\mathbb{Z}_5/\mathbb{Z}_2$ NISRs reduce to the trivial group --- see Eq.~\eqref{eq:Gr_odd} --- although they are exact at the tree level. This means that all the interaction terms that are forbidden in the classical Lagrangian but are generated radiatively are \emph{low-order zeros}. For instance, the mixing between $\phi_0$ and $\phi_1$ is not allowed at the tree level but is present at the one-loop level; see Fig.~\ref{fig:groupification} for a diagrammatic illustration of the generation of the mixing term $\phi_0 \phi_1$ at the one-loop order.

On the other hand, the spurion analysis in Section~\ref{sec:general_theory} yields 
\bea
\ell\left(\lambda_{1 2^2}\right) &=& [g^1], \quad \ell\left(\lambda_{1^2 2}\right) = [g^2], \quad \ell\left(\lambda_{0 1 2^2}\right) = [g^1], \quad \ell\left(\lambda_{0 1^2 2}\right) = [g^2], \nn\\
\ell\left(\lambda_{1 2^3}\right) &=& [g^1] [g^2] , \quad \ell\left(\lambda_{1^3 2}\right) = [g^1] [g^2]\ ,
\label{eq:label_example_Z5}
\eea
while all the other couplings are labeled by the identity. Correspondingly, these vertices carry nontrivial charges under the lifted group
\be
G_{\text{lift}} = \mathbb{Z}_2\times \mathbb{Z}_2 \; ,
\ee
where each $\mathbb{Z}_2$ factor indicates whether $[g^1]$ or $[g^2]$ appears in conjugate pairs. Specifically, the nontrivially charged vertices --- and hence the corresponding couplings --- are charged as 
\bea
c(\lambda_{1 2^2}) &=& (-1,1), \quad c(\lambda_{1^2 2}) = (1,-1), \quad c(\lambda_{0 1 2^2}) = (-1,1), \quad c(\lambda_{0 1^2 2}) = (1,-1), \nn\\
c(\lambda_{1 2^3}) &=& (-1,-1), \quad c(\lambda_{1^3 2}) = (-1,-1)\; ,
\label{eq:charge_example_Z5}
\eea
while all the other couplings are neutral (i.e., with the charge $(1,1)$) under the lifted group. The existence of these nontrivially-charged couplings implies that the $\mathbb{Z}_5/\mathbb{Z}_2$ NISRs can be interpreted as a special way of breaking the lifted group $\mathbb{Z}_2\times \mathbb{Z}_2$. Relatedly, the couplings in Eqs.~\eqref{eq:label_example_Z5} or~\eqref{eq:charge_example_Z5} are technically natural in the sense of 't Hooft~\cite{tHooft:1979rat}.

One can explicitly verify the consistency of Eqs.~\eqref{eq:spurion_tree} and~\eqref{eq:spurion_loop} in this concrete example. For instance, as shown in Fig.~\ref{fig:groupification}, the mixing between $\phi_0$ and $\phi_1$ is generated by radiative corrections at the one-loop order. According to our labeling scheme in Section~\ref{sec:general_theory}, the overall coupling of the $\phi_0\phi_1$ vertex is labeled as
\be
\ell\left(\lambda_{01}\right)=[g^1] \;,
\label{eq:label_example_Z5_loop}
\ee
which corresponds to the charge under the lifted group $c\left(\lambda_{01}\right) = (-1,1)$.
We find 
\be
\ell\left(\lambda_{01}\right) \prec \ell\left(\lambda_{1 2^2}\right) \ell\left(\lambda_{0 2^2}\right) \;.
\ee
Under the lifted group, this corresponds to 
\be
c\left(\lambda_{01}\right) = c\left(\lambda_{1 2^2}\right) c\left(\lambda_{0 2^2}\right) \;.
\ee
In this example, we see explicitly that the $\mathbb{Z}_5/\mathbb{Z}_2$ NISRs enforce a pattern of hierarchical couplings that cannot be explained using only the broken lifted group $G_{\text{lift}}$: the couplings $\lambda_{01}$ carries the same charge as $\lambda_{12^2}$ under $G_{\text{lift}}$, but only $\lambda_{12^2}$ is allowed at the tree level, as dictated by the $\mathbb{Z}_5/\mathbb{Z}_2$ fusion algebra. 

\begin{figure}[t]
\centering
\includegraphics[scale=0.2]{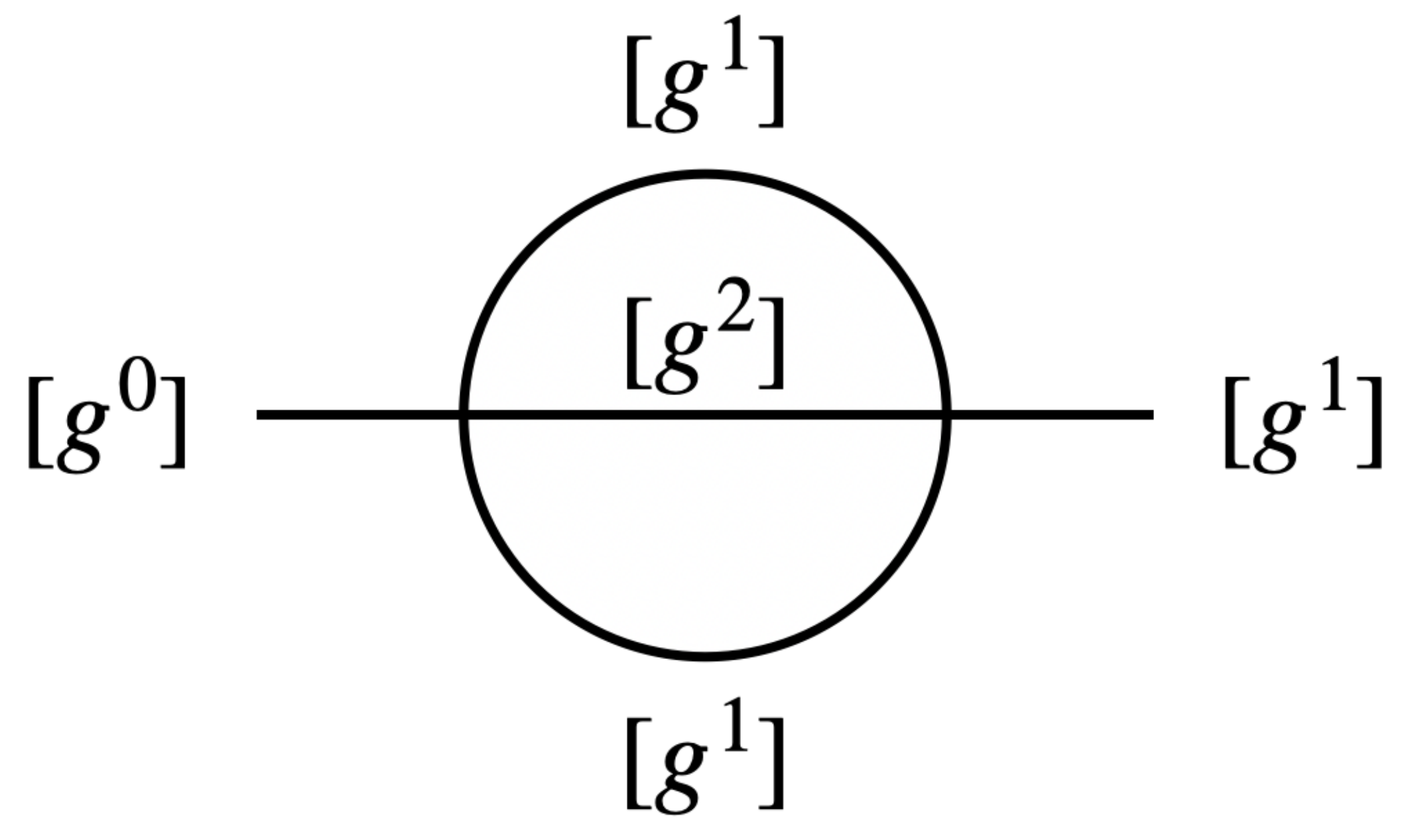}
\caption{Radiative generation of the $\phi_0\phi_1$ mixing term at the two-loop order. Here the particles are labeled by the elements in the $\mathbb{Z}_5/\mathbb{Z}_2$ fusion algebra; see Eq.~\eqref{eq:exp_particle_labels}. } 
\label{fig:Exp_2loop}
\end{figure}

Moreover, one may also consider the radiative generation of $\phi_0\phi_1$ at the two-loop order through the tree-level vertices $\phi_0 \phi_1^2 \phi_2$ and $\phi_1^3 \phi_2$; see Fig.~\ref{fig:Exp_2loop} for illustration. From Eqs.~\eqref{eq:label_example_Z5} and~\eqref{eq:label_example_Z5_loop}, one can easily see that 
\be
\ell\left(\lambda_{01}\right) \prec \ell\left(\lambda_{01^22}\right) \ell\left(\lambda_{1^32}\right)\;,
\ee
which again is consistent with the group law of the lifted group $\mathbb{Z}_2\times \mathbb{Z}_2$:
\be
c\left(\lambda_{01}\right) = c\left(\lambda_{01^22}\right) c\left(\lambda_{1^32}\right)\;.
\ee

Next, we consider the case where the $\mathbb{Z}_5/\mathbb{Z}_2$ fusion algebra is \emph{not} faithfully realized. As discussed in Section~\ref{sec:algebra_review}, particular care must be taken when applying the mechanism of loop-induced groupification in such situations. For example, consider a theory with only dynamical particles $\phi_0$ and $\phi_1$, whose classical Lagrangian includes the following interaction terms consistent with the $\mathbb{Z}_5/\mathbb{Z}_2$ NISRs, up to higher-dimensional operators:
\bea
\mathcal{L}_{\mathbb{Z}_5/\mathbb{Z}_2} (\phi_0, \phi_1) = \lambda_{0} \ \phi_0 + \lambda_{0^2} \ \phi_0^2 + \lambda_{1^2} \ \phi_1^2 + \lambda_{0^3} \ \phi_0^3 + \lambda_{0 1^2} \ \phi_0 \phi_1^2 + \lambda_{0^4} \ \phi_0^4  + \lambda_{0^2 1^2} \ \phi_0^2 \phi_1^2 + \lambda_{1^4} \ \phi_1^4 \;.\nn \\
\label{eq:Lag_Z5_2}
\eea
From the interaction terms in Eq.~\eqref{eq:Lag_Z5_2}, the mixing between $\phi_0$ and $\phi_1$ can \emph{never} be radiatively generated, as seen easily using spurion analysis. Although the $\mathbb{Z}_5/\mathbb{Z}_2$ NISRs would predict that the vanishing coupling $\lambda_{01}$ is only a low-order zero --- forbidden at the tree level but expected to be generated through radiative corrections via loop-induced groupification --- in this example the groupification mechanism is obstructed due to the nonfaithful realization of the $\mathbb{Z}_5/\mathbb{Z}_2$ fusion algebra.

One might view the theory of Eq.~\eqref{eq:Lag_Z5_2} as a low-energy effective theory of Eq.~\eqref{eq:Lag_Z5}, where the particle $\phi_2$ has a large mass and hence decouples at infrared. However, when integrating out $\phi_2$, extra terms involving unpaired $\phi_1$ --- e.g., $\phi_0\phi_1$ and $\phi_1^5$ etc --- must be added to Eq.~\eqref{eq:Lag_Z5_2} at the matching scale, such that the two Lagrangians yield the same physics at low energies.~\footnote{Here, we infer the existence of these additional terms purely from the perspective of groupification of NISRs, rather than from an explicit matching computation in effective theory.} Without these terms, Eq.~\eqref{eq:Lag_Z5_2} and Eq.~\eqref{eq:Lag_Z5} describe intrinsically distinct theories that cannot be related through renormalization-group (RG) flows. 

More broadly, one may conjecture the following:
\\[0.25cm]
\textit{Two Lagrangians constructed from distinct non-invertible fusion algebras cannot be related by RG flows if their all-order exact groups obtained from groupification are not the same.}
\\[0.4cm]
Indeed, in the absence of the terms involving unpaired $\phi_1$, the Lagrangian of Eq.~\eqref{eq:Lag_Z5_2} has an all-order exact $\mathbb{Z}_2$ group under which $\phi_1\to -\phi_1$~\footnote{Another view on the Lagrangian of Eq.~\eqref{eq:Lag_Z5_2} is that it is constructed based on the NISRs from $\mathbb{Z}_2$ orbifolding of a $\mathbb{Z}_2$ symmetry that is realized faithfully, where $\phi_0$ and $\phi_1$ are labeled by $[g^0]$ and $[g^1]$, respectively.}, whereas the Lagrangian of Eq.~\eqref{eq:Lag_Z5} has only the trivial group that remains exact at all loop orders --- see Eq.~\eqref{eq:Gr_odd}.
Hence, the two Lagrangians cannot be related by RG flows without the terms breaking the $\mathbb{Z}_2$ in Eq.~\eqref{eq:Lag_Z5_2}.

\section{Low-order and all-order zeros in Yukawa textures}
\label{sec:zeros}

We investigate the emergence of low-order and all-order zeros in Yukawa texture structures, inspired by the analysis of~\cite{Kobayashi:2024cvp}.
Specifically, we consider three types of textures based on $\mathbb{Z}_5/\mathbb{Z}_2$ and $\mathbb{Z}_6/\mathbb{Z}_2$ NISRs.
In the first case with $\mathbb{Z}_5/\mathbb{Z}_2$ NISRs, all the entries that vanish at the tree level become nonzero through higher-loop corrections.
In the second case with $\mathbb{Z}_6/\mathbb{Z}_2$ NISRs, all the zero entries are preserved to all loop orders in perturbation theory.
In the third case with $\mathbb{Z}_6/\mathbb{Z}_2$ NISRs, some entries remain zero to all loop orders, while others become nonzero with radiative corrections at loop order.

These examples demonstrate that our spurion analysis accurately captures the structure of Yukawa textures, identifying which zeros are protected and which can be generated radiatively.

\subsection{Low-order zeros in $\mathbb{Z}_5/\mathbb{Z}_2$}

The Higgs field is labeled by $[g^1]$, while the quark fields are labeled by $[g^0],~[g^1],~[g^2]$ corresponding to the three generations. This leads to the following Yukawa texture at the tree level:
\begin{align}
Y_{ij}\sim 
\begin{pmatrix}
0 & a & 0\\
b & 0 & c\\
0 & d & e
\end{pmatrix}\ ,
\end{align}
where $a,b,c,d, e$ denote nonzero independent complex parameters.

Our spurion analysis assigns nontrivial labels to some matrix elements as
\begin{align}
\label{eq:low_y_label}
\ell(Y_{ij})=
\begin{pmatrix}
[g^1] & [g^0] & [g^1][g^2]\\
[g^0] & [g^1] & [g^2]\\
[g^1][g^2] & [g^2] & [g^1]
\end{pmatrix}\ .
\end{align}
This indicates that all the texture zeros are low-order zeros. Through the element $Y_{33}$, which is labeled by $[g^1]$, all the texture zeros present at the tree level become nonzero through radiative corrections. Specifically, at the one-loop order, we find
\bea
\ell\left(Y_{22}\right) &\prec & \ell\left(Y_{23}\right) \ell\left(Y_{33}\right) \ell\left(Y_{32}\right), \\
\ell\left(Y_{13}\right) &\prec & \ell\left(Y_{12}\right) \ell\left(Y_{32}\right) \ell\left(Y_{33}\right), \\
\ell\left(Y_{31}\right) &\prec & \ell\left(Y_{33}\right) \ell\left(Y_{23}\right) \ell\left(Y_{21}\right).
\eea
At the two-loop order, we have 
\be
\ell\left(Y_{11}\right) \prec  \ell\left(Y_{12}\right) \ell\left(Y_{22}\right) \ell\left(Y_{21}\right),
\ee
where $Y_{22}$ is itself generated at the one-loop level.

\subsection{All-order zeros in $\mathbb{Z}_6/\mathbb{Z}_2$}

The Higgs field is labeled by $[g^1]$, while the quark fields are labeled as $[g^0],~[g^1],~[g^2]$ for the three generations,~\footnote{This labeling is unfaithful. One may alternatively consider $\mathbb{Z}_4/\mathbb{Z}_2$ NISRs, for which the labeling becomes faithful. However, this leads to the same conclusion regarding the all-order zeros.}
leading to the following Yukawa texture:
\begin{align}
\label{eq:all_zero_yukawa}
Y_{ij}\sim
\begin{pmatrix}
0 & a & 0\\
b & 0 & c\\
0 & d & 0
\end{pmatrix}\ ,
\end{align}
where $a$, $b$, $c$, and $d$ denote nonzero independent complex parameters.

According to our spurion analysis, the corresponding couplings are labeled as
\begin{align}
\ell(Y_{ij})=
\begin{pmatrix}
[g^1] & [g^0] & [g^1][g^2]\\
[g^0] & [g^1] & [g^2]\\
[g^1][g^2] & [g^2] & [g^1]
\end{pmatrix}\ .
\end{align}
This labeling is identical to that in~Eq.~\eqref{eq:low_y_label}.
However, in the present case, all nonzero entries are labeled by $[g^M]$ with even $M$, whereas all the zero entries contain $[g^1]$, corresponding to $[g^M]$ with odd $M$. According to the $\mathbb{Z}_6/\mathbb{Z}_2$ fusion algebra, $[g^1]$ is not contained in the fusion product of any $[g^M]$ with even $M$. Consequently, the spurion analysis indicates that none of the zeros can receive radiative corrections --- and hence become nonzero --- at any loop order in perturbation theory.

These all-order exact zeros can also be understood as a consequence of the $\mathbb{Z}_2$ symmetry arising from groupification, under which all nonzero entries carry the charge $+1$, while the zero entries carry the charge $-1$.~\footnote{
The determinant of the matrix in~Eq.~\eqref{eq:all_zero_yukawa} vanishes, indicating the presence of a chiral symmetry.
Introducing small nonzero entries in place of the zeros in~Eq.~\eqref{eq:all_zero_yukawa} would explicitly break this chiral symmetry by rendering the determinant nonzero.
This shows that the zeros in~Eq.~\eqref{eq:all_zero_yukawa} are all-order zeros.
}

\subsection{Low-order and all-order zeros in $\mathbb{Z}_6/\mathbb{Z}_2$}

The Higgs field is labeled by $[g^2]$, and the quark fields are labeled as $[g^1],~[g^2],~[g^3]$ corresponding to the three generations, which leads to the following Yukawa texture:
\begin{align}
\label{eq:all_zero_yukawa_2}
Y_{ij}\sim
\begin{pmatrix}
a & 0 & b\\
0 & c & 0\\
d & 0 & 0
\end{pmatrix}\ .
\end{align}
From our spurion analysis, the couplings are labeled as
\begin{align}
\ell(Y_{ij})=
\begin{pmatrix}
[g^2] & [g^1] & [g^1][g^2][g^3]\\
[g^1] & [g^2] & [g^3]\\
[g^1][g^2][g^3] & [g^3] & [g^2]
\end{pmatrix}\ .
\end{align}
Our spurion analysis indicates that the zeros at the entries $Y_{12}, Y_{21}, Y_{23}, Y_{32}$ are exact to all loop orders in perturbation theory, while the zero of $Y_{33}$ is a low-order zero --- i.e., radiative corrections can generate a nonzero value for $Y_{33}$. Indeed, at one-loop order, we have
\be
\ell\left(Y_{33}\right) \prec  \ell\left(Y_{31}\right) \ell\left(Y_{11}\right) \ell\left(Y_{13}\right).
\ee
Therefore, the Yukawa texture, including radiative corrections, takes the form
\begin{align}
Y_{ij}^{\rm loop}\sim
\begin{pmatrix}
a & 0 & b\\
0 & c & 0\\
d & 0 & \epsilon
\end{pmatrix}\ .
\end{align}

\section{Conclusion}
\label{sec:conclusion}

While selection rules derived from group theory are standard tools in particle physics, those emerging from non-invertible fusion algebras remain comparatively unexplored --- both in terms of their mathematical structure and their implications for perturbation theory and model building in particle physics. Recently, however, this direction has seen a revival~\cite{Kaidi:2024wio, Heckman:2024obe, Kobayashi:2024yqq, Kobayashi:2024cvp, Funakoshi:2024uvy, Kobayashi:2025znw, Suzuki:2025oov, Suzuki:2025bxg, Liang:2025dkm, Kobayashi:2025ldi, Kobayashi:2025cwx, Nomura:2025sod, Kobayashi:2025lar, Dong:2025jra, Nomura:2025yoa, Chen:2025awz, Okada:2025kfm, Kobayashi:2025thd, Jangid:2025krp, Kobayashi:2025ocp, Kobayashi:2025rpx, Jiang:2025psz, Nomura:2025tvz, Jangid:2025thp}, revealing that such non-invertible selection rules (NISRs) can even arise in some of the simplest extensions of the Standard Model. Remarkably, they predict distinctive hierarchical patterns of couplings that cannot be explained by ordinary symmetries, demonstrating that NISRs are far from exotic~\cite{Suzuki:2025oov} and may, in fact, be essential ingredients of realistic theories.~\footnote{Another application similar to the ones in~\cite{Suzuki:2025oov} is to explain the suppression of the kinetic mixing between dark photon and the SM photon (see e.g.~\cite{Banerjee:2019asa, Dienes:1996zr, Holdom:1985ag}) using the Fibonacci fusion rules, where the mixing is only radiative generated through groupification. We might revisit this subject in the future.}

To enable systematic applications of NISRs, and motivated in part by the models analyzed in~\cite{Suzuki:2025oov}, we previously initiated a program~\cite{Suzuki:2025bxg} aimed at reinterpreting the mechanism of ``loop-induced groupification''~\cite{Kaidi:2024wio} through the lens of spurion analysis, focusing on NISRs derived from near-group fusion algebras~\cite{Evans:2012ta}. See e.g.~\cite{Suzuki:2025oov, Kobayashi:2025cwx, Nomura:2025sod, Chen:2025awz, Jangid:2025krp, Jangid:2025thp} on the use of near-group fusion algebras in particle physics models. In the present work, we extend the framework of spurion analysis to NISRs arising from $\mathbb{Z}_M/\mathbb{Z}_2$ orbifolding, motivated by recent studies~\cite{Kobayashi:2024yqq, Kobayashi:2024cvp, Funakoshi:2024uvy, Kobayashi:2025znw, Liang:2025dkm, Kobayashi:2025ldi, Kobayashi:2025lar, Nomura:2025yoa, Okada:2025kfm, Kobayashi:2025thd, Kobayashi:2025rpx, Jiang:2025psz, Nomura:2025tvz} in the literature. The main results of this work can be summarized as follows:
\begin{itemize}
\item Although the $\mathbb{Z}_M/\mathbb{Z}_2$ NISRs are exact at the tree-level, some of the couplings need to be promoted to non-identity elements in the spurion analysis. These couplings are the sources of radiative breaking of the NISRs. The labeling scheme developed in Section~\ref{sec:spurion-analysis} enables us to systematically track such couplings in all the scattering processes at both tree and loop levels in perturbation theory. 

This provides an alternative interpretation of the mechanism of ``loop-induced groupification'' for the all-order exact group. Relatedly, the spurion analysis offers a transparent understanding of the low-order versus all-order zeros enforced by the $\mathbb{Z}_M/\mathbb{Z}_2$ NISRs, as illustrated through many examples in Sections~\ref{sec:low_order_zeros_decoupling} and~\ref{sec:zeros}.

\item The lifted group $G_{\text{lift}}$ --- see Eq.~\eqref{eq:lifted_G} --- provides a complementary perspective in the spurion analysis and ensures the validity of the results in Section~\ref{sec:spurion-analysis}. Specifically, the $\mathbb{Z}_M/\mathbb{Z}_2$ NISRs can be viewed as a special way of breaking the lifted group, where the couplings labeled by non-identity elements of the fusion algebra correspondingly carry nontrivial charges under $G_{\text{lift}}$, guaranteeing their technical naturalness in the sense of 't Hooft. Nevertheless, the $\mathbb{Z}_M/\mathbb{Z}_2$ NISRs predict hierarchical structures among couplings that cannot be explained using only the broken $G_{\text{lift}}$. See Sections~\ref{sec:spurion-analysis} and~\ref{sec:low_order_zeros_decoupling} for further details.

\item In Section~\ref{sec:algebra_review}, we provide some original remarks on the situations where the standard mechanism of ``loop-induced groupification'' is obstructed. In Section~\ref{sec:low_order_zeros_decoupling}, we initiate the study of the cases where the fusion algebra is not faithfully realized, and we formulate a conjecture on the related aspects on particle decoupling and effective theories. A more systematic understanding is left to future work.
\end{itemize}

Here we outline several possible future directions.
For both the near-group fusion algebras studied in~\cite{Suzuki:2025bxg} and the $\mathbb{Z}_M/\mathbb{Z}_2$ analyzed in the present paper, all the non-invertible elements are self-conjugate, hence the corresponding lifted group for each such element is $\mathbb{Z}_2$. A natural next step is to extend the framework of spurion analysis to fusion algebras where the non-invertible elements are not self-conjugate.
It is also promising to explore the interplay between NISRs with supersymmetry, where the superpotential obeys the perturbative non-renormalization theorem~\cite{Grisaru:1979wc, Seiberg:1993vc}, and with locality in the theory space (i.e., the deconstructed extra dimension)~\cite{Arkani-Hamed:2001kyx}. Those scenarios are related to the situations discussed in Section~\ref{sec:algebra_review}, where the vanilla argument of loop-induced groupification does not apply. Some of these directions will be pursued elsewhere.

\section{Acknowledgment}
We thank Hao Y. Zhang for related collaborations.
We thank Joan Elias Miro for interesting comments.
M.S. is supported by the MUR projects 2017L5W2PT.
M.S. also acknowledges the European Union - NextGenerationEU, in the framework of the PRIN Project “Charting unexplored avenues in Dark Matter” (20224JR28W).
The work of L.X.X. is partially supported by ``Exotic High Energy Phenomenology" (X-HEP), a project funded by the European Union - Grant Agreement n.101039756. Funded by the European Union. Views and opinions expressed are however those of the author(s) only and do not necessarily reflect those of the European Union or the ERC Executive Agency (ERCEA). Neither the European Union nor the granting authority can be held responsible for them. This project is also supported by the Munich Institute for Astro-, Particle and BioPhysics (MIAPbP) which is funded by the Deutsche Forschungsgemeinschaft (DFG, German Research Foundation) under Germany's Excellence Strategy – EXC-2094 – 390783311.

\begin{appendix}
\section{Labeling with basis elements with $M=5,6,7$}
\label{app:A}

In Section~\ref{sec:spurion-analysis}, we propose a general labeling scheme for performing the spurion analysis for $\mathbb{Z}_M/\mathbb{Z}_2$ NISRs and verify its consistency. The key idea is the use of a lifted group --- see Eq.~\eqref{eq:lifted_G} --- which naturally leads to composite labeling, where a coupling is labeled by the fusion product of basis elements rather than by a single basis element.

A natural question then arises: 
\\[0.25cm]
\textit{Can we find an alternative labeling scheme in which each coupling is associated with only a single basis element? }
\\[0.4cm]
In this approach, we continue to apply the same rule for self-conjugate pairs (which is to replace them by the identity), but for the couplings we select one basis element from the corresponding fusion product instead of retaining the entire product. Such a choice, however, cannot be made arbitrarily --- it must be implemented consistently so that Eqs.~\eqref{eq:spurion_tree} and~\eqref{eq:spurion_loop} remain valid. 

We examine three examples of the $\mathbb{Z}_M/\mathbb{Z}_2$ fusion algebra with $M=5,6,7$ and demonstrate that a consistent assignment of a single basis element to each coupling is indeed possible.

\subsection{$M=5$}

When $M=5$, the lifted group is 
\be
G_{\text{lift}} =\mathbb{Z}_2\times \mathbb{Z}_2\;,
\ee
where each $\mathbb{Z}_2$ factor indicates whether $[g^1]$ or $[g^2]$ --- denoting external dynamical particles in a scattering amplitude --- appear in pairs. All possible amplitudes can be classified into four types according to their charges under $G_{\text{lift}}$:
\be
\mathcal{M}_{(1,1)},~\mathcal{M}_{(1,-1)},~\mathcal{M}_{(-1,1)},~\mathcal{M}_{(-1,-1)}\ .
\ee
Following the discussion in Section~\ref{sec:general_theory}, these amplitudes $\mathcal{M}$'s --- and therefore their corresponding couplings $\lambda_{\mathcal{M}}$'s --- are labeled as
\begin{align}
& \mathcal{M}_{(1,1)} :~ [g^0] = \mathbbm{1}\ ,\\
& \mathcal{M}_{(1,-1)} :~[g^2]\ ,\\
& \mathcal{M}_{(-1,1)} :~ [g^1]\ ,\\
& \mathcal{M}_{(-1,-1)} :~ [g^1][g^2]\ .
\end{align}
It is straightforward to verify that the labeling above satisfies Eqs.~\eqref{eq:spurion_tree} and~\eqref{eq:spurion_loop}, e.g., 
\bea
\ell\left(\lambda_{\mathcal{M}_{(-1,-1)}}\right) &\prec & \ell\left(\lambda_{\mathcal{M}_{(1,-1)}}\right) \ell\left(\lambda_{\mathcal{M}_{(-1,1)}}\right), \label{eq:label_M5_1}\\
\ell\left(\lambda_{\mathcal{M}_{(1,-1)}}\right) &\prec & \ell\left(\lambda_{\mathcal{M}_{(-1,1)}}\right) \ell\left(\lambda_{\mathcal{M}_{(-1,-1)}}\right), \label{eq:label_M5_2} \\
\ell\left(\lambda_{\mathcal{M}_{(-1,1)}}\right) &\prec & \ell\left(\lambda_{\mathcal{M}_{(1,-1)}}\right) \ell\left(\lambda_{\mathcal{M}_{(-1,-1)}}\right). \label{eq:label_M5_3}
\eea

Next, we select one basis element from the fusion product $[g^1][g^2]$ to label $\lambda_{\mathcal{M}_{(-1,-1)}}$. For $M=5$, the fusion rule reads 
\[
[g^1][g^2] = [g^1] \oplus [g^2]\ ,
\]
so either $[g^1]$ or $[g^2]$ can be chosen. 
It turns out that both options are consistent with Eqs.~\eqref{eq:label_M5_1}-\eqref{eq:label_M5_3}, leading to the following labeling scheme using only the basis elements:
\begin{align}
   & \ell\!\left(\lambda_{\mathcal{M}_{(1,1)}}\right) = \mathbbm{1}\ ,\\
   & \ell\!\left(\lambda_{\mathcal{M}_{(1,-1)}}\right) = [g^2]\ ,\\
   & \ell\!\left(\lambda_{\mathcal{M}_{(-1,1)}}\right) = [g^1]\ ,\\
   & \ell\!\left(\lambda_{\mathcal{M}_{(-1,-1)}}\right) = [g^1]~\text{or}~[g^2]\ .
\end{align}

\subsection{$M=6$}

For $M=6$, we have three nontrivial basis elements, corresponding to the lifted group
\be
G_{\text{lift}} =\mathbb{Z}_2\times \mathbb{Z}_2 \times \mathbb{Z}_2\;.
\ee
We therefore obtain $2^3 = 8$ types of amplitudes, classified by their charges under $G_{\text{lift}}$.  Following the discussion in Section~\ref{sec:general_theory}, the amplitudes and their corresponding couplings are labeled as
\begin{align}
   & \mathcal{M}_{(1,1,1)}:~ \mathbbm{1}\ ,\\
   & \mathcal{M}_{(-1,1,1)}:~ [g^1]\ ,\\
   & \mathcal{M}_{(1,-1,1)}:~ [g^2]\ ,\\
   & \mathcal{M}_{(1,1,-1)}:~ [g^3]\ ,\\
   & \mathcal{M}_{(-1,-1,1)}:~ [g^1][g^2] = [g^1] \oplus [g^3]\ ,\\
   & \mathcal{M}_{(-1,1,-1)}:~ [g^1][g^3] = [g^2]\ ,\\
   & \mathcal{M}_{(1,-1,-1)}:~ [g^2][g^3] = [g^1]\ ,\\
   & \mathcal{M}_{(-1,-1,-1)}:~ [g^1][g^2][g^3] = \mathbbm{1} \oplus [g^2]\ .
\end{align}

Next, we select the specific basis elements for the couplings $\lambda_{\mathcal{M}_{(-1,-1,1)}}$ and $\lambda_{\mathcal{M}_{(-1,-1,-1)}}$. Again, the requirement is the consistency with Eqs.~\eqref{eq:spurion_tree} and~\eqref{eq:spurion_loop}. 
We find two consistent labeling choices:
\begin{align}
   & \ell\!\left(\lambda_{\mathcal{M}_{(-1,-1,1)}}\right) = [g^1]\ ,~
   \ell\!\left(\lambda_{\mathcal{M}_{(-1,-1,-1)}}\right) = [g^2]\ ,
\end{align}
or
\begin{align}
   & \ell\!\left(\lambda_{\mathcal{M}_{(-1,-1,1)}}\right) = [g^3]\ ,~
   \ell\!\left(\lambda_{\mathcal{M}_{(-1,-1,-1)}}\right) = \mathbbm{1}\ .
\end{align}
This example illustrates that the choice cannot be made arbitrarily. 

\subsection{$M=7$}

For $M=7$, we have three nontrivial basis elements with the lift group
\be
G_{\text{lift}} =\mathbb{Z}_2\times \mathbb{Z}_2 \times \mathbb{Z}_2\;,
\ee
and thus we obtain eight types of amplitudes with the labeling:
\begin{align}
   & \mathcal{M}_{(1,1,1)}:~ \mathbbm{1}\ ,\\
   & \mathcal{M}_{(-1,1,1)}:~ [g^1]\ ,\\
   & \mathcal{M}_{(1,-1,1)}:~ [g^2]\ ,\\
   & \mathcal{M}_{(1,1,-1)}:~ [g^3]\ ,\\
   & \mathcal{M}_{(-1,-1,1)}:~ [g^1][g^2] = [g^1] \oplus [g^3]\ ,\\
   & \mathcal{M}_{(-1,1,-1)}:~ [g^1][g^3] = [g^2] \oplus [g^3]\ ,\\
   & \mathcal{M}_{(1,-1,-1)}:~ [g^2][g^3] = [g^1] \oplus [g^2]\ ,\\
   & \mathcal{M}_{(-1,-1,-1)}:~ [g^1][g^2][g^3] = \mathbbm{1} \oplus [g^1] \oplus [g^2] \oplus [g^3]\ .
\end{align}
We find a single consistent labeling given by
\begin{align}
  & \ell\!\left(\lambda_{\mathcal{M}_{(-1,-1,1)}}\right) = [g^3]\ ,~~
    \ell\!\left(\lambda_{\mathcal{M}_{(-1,1,-1)}}\right) = [g^2]\ ,~~
    \ell\!\left(\lambda_{\mathcal{M}_{(1,-1,-1)}}\right) = [g^1]\ ,~~
    \ell\!\left(\lambda_{\mathcal{M}_{(-1,-1,-1)}}\right) = \mathbbm{1}\ .
\end{align}
This example illustrates that the choice of basis elements cannot be made arbitrarily.

\end{appendix}

\bibliography{NoninvertSpurion.bib}

\end{document}